\documentclass[aps,prd,superscriptaddress,twoside,twocolumn,nofootinbib,10pt,%
showpacs]{revtex4-1}

\usepackage{amsmath,amssymb}
\usepackage{dcolumn}
\usepackage{ulem} 
\usepackage[usenames]{color}

\allowdisplaybreaks

\begin{document}

\title{Doubly heavy baryons with chiral partner structure}

\author{Yong-Liang Ma}
\email{yongliangma@jlu.edu.cn}
\affiliation{Center of Theoretical Physics and College of Physics, Jilin University, Changchun, 130012, China}

\author{Masayasu Harada}
\email{harada@hken.phys.nagoya-u.ac.jp}
\affiliation{Department of Physics,  Nagoya University, Nagoya, 464-8602, Japan}

\date{\today}
\begin{abstract}
The spectrum and dominant strong decay properties of the doubly heavy baryons are revisited by using a chiral effective model with chiral partner structure. By regarding the doubly heavy baryons in the ground states and light angular momentum $j_l = 1/2$ sector of the first orbitally excited states as chiral partners to each other, we estimate the mass splitting of the chiral partners which arises from the spontaneous breaking of chiral symmetry to be about $430$~MeV for baryons including an unflavored light quark and about $350$~MeV for that including a strange quark. We point out that, similar to the heavy-light meson sector, the intermultiplet decay from a baryon with negative parity to its chiral partner and a pion is determined by the mass splitting throught the generalized Goldberger-Treiman relation. Furthermore, the isospin violating decay of $\Omega_{cc}$ baryon, $((1/2)^-, (3/2)^-)_s \to ((1/2)^+, (3/2)^+)_s + \pi^0$ through the $\eta$-$\pi^0$ mixing is the dominant decay channel of the doubly heavy baryons including a strange quark.
\end{abstract}

\maketitle

The heavy hadron spectroscopy has drown extensive attention since last decade because of the observation of the large amount of heavy hadrons in particle colliders. It is reasonable to expect that more and heavier resonances, such as the doubly heavy baryons (DHBs) concerned in this work, could be observed by the ongoing and future scientific facilities such as LHCb and Belle II.

The existence of DHBs is an immediate prediction of QCD. The theoretical discussion of these baryons has been done for a long time~\cite{Moinester:1995fk,Brodsky:2011zs,Sun:2014aya}. Meanwhile, several experimental efforts have been made to detect such states and the positive results from SELEX shows that the mass of the doubly charmed baryons $\Xi_{cc}^+$ is about $3520$~MeV~\cite{Agashe:2014kda}. In this paper, we investigate some properties of the DHBs based on the chiral partner structure. Since there is only one light quark in a DHB, unlike the light baryons, its chiral behaviour is quite simple.

The chiral partner structure of hadrons including heavy quark has been studied by several groups. In the heavy-light meson sector, the pioneer idea was proposed by Nowak {\it et al.} in Ref.~\cite{Nowak:1992um} and then Bardeen and Hill~\cite{Bardeen:1993ae}. In this picture, the heavy-light meson doublets in the heavy quark limit with quantum numbers ($1^-, 0^-$) and ($1^+, 0^+$) are regarded as chiral partners to each other and the mass splitting
of them is induced by the dynamical breaking of the chiral
symmetry so that the magnitude is about the constituent
quark mass. This
was confirmed by the spectrum of the relevant particles:
$m_{D_0^{\ast}} - m_D \simeq m_{D_1}- m_{D^{\ast}} \simeq 450~$MeV
is at the same order of $m_{D_{s0}}(2317) - m_{D_s} \simeq
m_{D_{s1}}(2460)- m_{D_{s}^{\ast}} \simeq 350~$MeV (see, e.g., Refs.~\cite{Bardeen:2003kt,Nowak:2003ra,Nowak:2004jg}). In the sector of heavy baryons including a heavy quark, the chiral partner structure are mainly accessed based on the bound state approach (see, e.g., Ref.~\cite{Harada:2012dm} and references therein). In this sector, there are some disagreements about the chiral partner structure. Normally, the chiral partner of the ground state baryon $\Lambda_c(2268)(J^P = \frac{1}{2}^+)$ is regarded as $\Lambda_c(2595)$ with $J^P = \frac{1}{2}^-$. However, a recent analysis with including the lowest lying vector mesons $\rho$ and $\omega$ meson effects shows that the chiral
partner of $\Lambda_c(\frac{1}{2}^+,2286)$
is the $\Lambda_c(\frac{1}{2}^-,\frac{3}{2}^-)$ heavy quark doublet with a mass of about
$3.1$\,GeV~\cite{Harada:2012dm}. This disagreement might arise from the fact that, in contrast to the heavy-light meson sector and also the DHBs considered in this work, in the heavy baryons including one heavy quark, there are two light quarks so their chiral properties are not so simple.

Schematically, the quark contents of a DHB can be written as $QQq$ with $Q$ and $q$ being the heavy quark and light quark constituents, respectively. Since the DHB is a colorless object, the two heavy quarks in it form an anti-color triplet~\cite{Bardeen:2003kt}. Because the heavy quarks in the DHB have a large mass, it takes a much larger energy to orbitally excite the heavy constituent quark than to orbitally excite the light one, it is reasonable to regard the two constituent quarks as an static compact object without orbital excitation and denote the constituent of the DHB as $\bar{\bm{\Phi}}q$ with $\bar{\Phi}$ being the heavy quark component which should be a bosonic quantity. With such an intuitive picture in mind, one can define the chiral partner structure similar to that in the heavy-light meson sector~\cite{Nowak:1992um,Bardeen:1993ae}.

Since two heavy quarks in a DHB is antisymmetric in color space, they should have the total spin $J_Q = 1$ in $s$-wave and, therefore DHBs in the ground states can form a heavy quark doublet $D_\textbf{Q}^\mu$ whose components have quantum numbers $J^P = \frac{1}{2}^+,\frac{3}{2}^+$. For the first orbital excitation with relative angular momentum between the light quark and heavy quark source $l = 1$, the light angular momentum could be $j_l = \frac{1}{2}, \frac{3}{2}$. Combing the $j_l = \frac{1}{2}$ one can form another heavy quark doublet $N_\textbf{Q}^\mu$ with quantum numbers $J^P = \frac{1}{2}^-,\frac{3}{2}^-$. We regard the doublets $D_\textbf{Q}^\mu$ and $N_\textbf{Q}^\mu$ as chiral partners to each other and DHBs constructed from $j_l = \frac{3}{2}$ can be regarded as chiral partners of some states from $l =2$ baryons~\cite{Bardeen:2003kt}.

Similar to the heavy-light meson case, since there is only one light quark in a DHB, we can write the DHB doublets $D_\textbf{Q}^\mu$ and $N_\textbf{Q}^\mu$ in the chiral basis by introducing the fields $D_{\textbf{Q};L,R}^\mu$ which at the quark level are schematically written as $D_{\textbf{Q};L,R}^\mu \sim \bar{\bm{\Phi}}^\mu q_{L, R}$. Since the heavy quark component of the DHB is a boson, $D_{\textbf{Q};L,R}^\mu$ should be Lorentz spinors and, under chiral transformation, transform as
\begin{eqnarray}
D_{\textbf{Q};L,R}^\mu & \to & g_{L,R} D_{\textbf{Q};L,R}^\mu ,
\label{eq:chiralDLR}
\end{eqnarray}
where $g_{L,R} \in SU(3)_{L,R}$. In terms of the $D_{\textbf{Q}}$ and $N_{\textbf{Q}}$, we can write
\begin{eqnarray}
D_{\textbf{Q};L}^\mu & = & \frac{1}{\sqrt{2}}\left(D_{\textbf{Q}}^\mu - i N_{\textbf{Q}}^\mu \right) , \nonumber\\
D_{\textbf{Q};R}^\mu & = & \frac{1}{\sqrt{2}}\left(D_{\textbf{Q}}^\mu + i N_{\textbf{Q}}^\mu \right),
\label{eq:ChiralPhys}
\end{eqnarray}
which transform as $D_{\textbf{Q};L, R}^\mu \leftrightarrow \gamma_0 D_{\textbf{Q};\mu;R, L}$ under parity transformation and satisfy $v\hspace{-0.17cm}\slash D_{\textbf{Q};L,R}^\mu = D_{\textbf{Q};L,R}^\mu$ and $v_\mu D_{\textbf{Q};L,R}^\mu = 0$ for preserving the heavy quark symmetry and keeping the transversality. And, for later convenience, following the procedure given in Ref.~\cite{Falk:1991nq}, we write the DHB doublets $D_\textbf{Q}^\mu$ and $N_\textbf{Q}^\mu$ in terms of the physical states as
\begin{eqnarray}
D_{\textbf{Q}}^\mu & = & \frac{1 + v \hspace{-0.17cm}\slash}{2}\Psi_{QQ}^\mu + \sqrt{\frac{1}{3}}\left(\gamma^\mu + v^\mu\right)\gamma^5\frac{1 + v \hspace{-0.17cm}\slash}{2}\Psi_{QQ} , \nonumber\\
N_{\textbf{Q}}^\mu & = & \frac{1 + v \hspace{-0.17cm}\slash}{2}\Psi_{QQ}^{\prime \mu} + \sqrt{\frac{1}{3}}\left(\gamma^\mu + v^\mu\right)\gamma^5\frac{1 + v \hspace{-0.17cm}\slash}{2}\Psi_{QQ}^{\ast} ,
\end{eqnarray}
which is the same as that for the heavy baryons including one heavy quark~\cite{Georgi:1990cx,Cho:1992gg} and $\Psi_{QQ}^{(\prime)\mu}$ is the spin-$\frac{3}{2}$ Rarita-Schwinger field. One can easily check that these spinors satisfy $v\hspace{-0.17cm}\slash D_{\textbf{Q}}^\mu = D_{\textbf{Q}}^\mu$ and $v\hspace{-0.17cm}\slash N_{\textbf{Q}}^\mu = N_{\textbf{Q}}^\mu$. We have imposed the intrinsic parity behaviour
\begin{eqnarray}
{\rm P} &:&~\Psi_{QQ}^\mu \to{}- \gamma_0 \Psi_{QQ,\mu},~~ \Psi_{QQ} \to \gamma_0 \Psi_{QQ} , \nonumber\\
& & ~ \Psi_{QQ}^{\prime \mu} \to \gamma_0\Psi_{QQ, \mu}^{\prime},~~ \Psi_{QQ}^{\ast} \to {} - \gamma_0\Psi_{QQ}^{\ast}.
\end{eqnarray}
When the heavy quark in the DHB is $c$ quark and the light quark is either of $u, d$ and $s$ quarks, the DHB field, for example $\Psi_{QQ}$ stands for $\Xi_{cc}^{++}, \Xi_{cc}^+$ and $\Omega_{cc}^+$, respectively.

Now, we are in the position to construct the chiral effective theory of DHBs in the chiral basis. We note that the quark-diquark symmetry~\cite{Savage:1990di} relates the doubly heavy baryons with the heavy mesons having the same Brown muck~\cite{Hu:2005gf}. For relating the parameters based on the quark-diquark symmetry, we first write an effective Lagrangian for the heavy-light mesons with the chiral partner structure by introducing chiral fields $\mathcal{H}_{L,R}$~\cite{Nowak:1992um,Bardeen:1993ae}. These chiral fields relate to the heavy-light meson doublets $H$ and $G$ with quantum numbers $(0^-,1^-)$ and $(0^+,1^+)$, respectively, through
\begin{eqnarray}
\mathcal{H}_R = \frac{1}{\sqrt{2}}\left[ G - i H \gamma_5 \right] \ ,\ \ \mathcal{H}_L = \frac{1}{\sqrt{2}}\left[ G + i H \gamma_5 \right]\ ,
\label{eq:HLRGH}
\end{eqnarray}
where $G$ and $H$ are heavy-light meson fields with the positive and negative parity, respectively. In terms of the physical states, they are expressed as
\begin{eqnarray}
H & = & \frac{1 + v\hspace{-0.17cm}\slash }{2} \left[ D^{\ast\mu}\gamma_\mu +  i D \gamma_5 \right] \ , \notag\\
G & = & \frac{1 + v\hspace{-0.17cm}\slash }{2} \left[{} - D_1^{\prime\mu} \gamma_\mu \gamma_5 + D_0^\ast \right] \ .
\label{eq:HGphys}
\end{eqnarray}
It should be noticed that, since in the heavy-light meson fields, the heavy component is a heavy quark and the light component is a light antiquark, not the chiral fields $\mathcal{H}_{L,R}$ but their conjugates $\bar{\mathcal{H}}_{L,R} \equiv \gamma_0 \mathcal{H}_{L,R} \gamma_0$ transform as the chiral quark fields $q_{L,R}$ under chiral transformation, i.e., the same as Eq.~\eqref{eq:chiralDLR}. Here, we consider only the terms which survive in the heavy quark limit and including the terms up till one derivative. For the light mesons, we consider the chiral field $M$ which transforms as $M \to g_L M g_R^\dagger$ under chiral transformation. The effective Lagrangian is written as~\cite{Harada:2012km,Suenaga:2014sga}
\begin{eqnarray}
{\mathcal L}_{\rm M} & = & {\rm tr}\left[\mathcal{H}_L(iv\cdot\partial)\bar{\mathcal{H}}_L] + {\rm tr}[\mathcal{H}_R(iv\cdot\partial)\bar{\mathcal{H}}_R\right] \nonumber\\
& &{} - \Delta \mbox{tr} \left[ \mathcal{H}_L \bar{\mathcal{H}}_L + \mathcal{H}_R  \bar{\mathcal{H}}_R\right] \nonumber\\
& &{} - \frac{1}{2}\, g_\pi \mbox{tr}
\left[\mathcal{H}_L M \bar{\mathcal{H}}_R + \mathcal{H}_R M^{\dagger} \bar{\mathcal{H}}_L\right] \nonumber\\
& &{} + i \frac{g_{A}}{f_\pi}\mbox{tr}
\left[\mathcal{H}_L\gamma_5\gamma^{\mu}\partial_{\mu} M \bar{\mathcal{H}}_R
- \mathcal{H}_R\gamma_5\gamma^{\mu}\partial_{\mu}M^\dag \bar{\mathcal{H}}_L\right] \ , \label{pionlagrangian}
\end{eqnarray}
where $\Delta$ provides the mass shift to both $G$ and $H$ in the same direction. After a suitable choice of the potential sector of the light meson Lagrangian which will not be specified here, one can realize the chiral symmetry in the Nambu-Goldstone phase. In such a case, after the spontaneous breaking of the chiral symmetry, the meson field $M$ can be replaced by $\bar{M} + \tilde{M}$ with $\bar{M} = {\rm diag} (v,v, v_3)$ being the vacuum expectation value of the chiral field in the isospin limit, which corresponds to the quark condensate, and $\tilde{M}$ being the fluctuation fields. Then, this $g_\pi$ term provides the mass difference between $G$ and $H$ as
\begin{equation}
\Delta M_i = m_{G,i} - m_{H,i} = g_\pi  \bar{M}_{ii} \ ,
\label{massdif:meson}
\end{equation}
where the sub-indices $i$ stand for the light flavor with $i = 1,2$ and $3$ being $u, d$ and $s$ quark, respectively.  Here we use $v = f_\pi =$92.4\,MeV, so that we obtain $g_\pi = 4.65$ from $\Delta M_{u,d} = 430\,$MeV.
Note that the $g_\pi$ term also gives the interaction for the pionic transition between $G$ and $H$.  The relation between these two quantities are known as generalized Goldberger-Treiman relation~\cite{Nowak:1992um,Bardeen:1993ae}.
On the other hand, the $g_A$ term gives the interaction of the pionic transition within $G$ or $H$.  The value of $g_A$ is determined from the experimental value of $D^\ast \to D + \pi$ decay as $g_A = 0.56$~(see e.g. \cite{Harada:2012km}).

Now, let us consider the effective Lagrangian for the doubly heavy baryons.  As we stated above, the quark-diquark symemtry relates the Lagrangian to the above Lagrangian for the heavy mesons. The resultant effective Lagrangian is expressed as
\begin{eqnarray}
{\cal L}_{\rm B} & = & \bar{D}_{\textbf{Q};L}^\mu i v\cdot \partial D_{\textbf{Q};\mu;L} + \bar{D}_{\textbf{Q};R}^\mu i v\cdot \partial D_{\textbf{Q};\mu;R} \nonumber\\
& &{} - \Delta\left( \bar{D}_{\textbf{Q};L}^\mu D_{\textbf{Q};\mu;L} + \bar{D}_{\textbf{Q};R}^\mu D_{\textbf{Q};\mu;R} \right) \nonumber\\
& &{} - \frac{1}{2}g_\pi\left( \bar{D}_{\textbf{Q};L}^\mu M D_{\textbf{Q};\mu; R} + \bar{D}_{\textbf{Q};R}^\mu M^\dagger D_{\textbf{Q};\mu;L} \right) \nonumber\\
& &{} + \frac{ig_A}{f_\pi}\left[ \bar{D}_{\textbf{Q};L}^\mu\gamma_5\gamma^\nu \partial_\nu M D_{\textbf{Q};\mu;R} \right. \nonumber\\
& & \left. \qquad\quad\;\; {} + \bar{D}_{\textbf{Q};R}^\mu\gamma_5\gamma^\nu \partial_\nu M^\dagger  D_{\textbf{Q};\mu;L}\right] .
\label{eq:EffecL}
\end{eqnarray}
By substituting \eqref{eq:ChiralPhys} into the Lagrangian \eqref{eq:EffecL} and considering the spontaneous chiral symmetry breaking, one obtains the Lagrangian
\begin{widetext}
\begin{eqnarray}
{\cal L}_{\rm B} & = & \bar{D}_{\textbf{Q}}^\mu i v\cdot \partial D_{\textbf{Q};\mu} + \bar{N}_{\textbf{Q}}^\mu i v\cdot \partial N_{\textbf{Q};\mu} - \Delta\left( \bar{D}_{\textbf{Q}}^\mu D_{\textbf{Q};\mu} + \bar{N}_{\textbf{Q}}^\mu N_{\textbf{Q};\mu} \right)
- \frac{1}{2} g_\pi\left( \bar{D}_{\textbf{Q}}^\mu \bar{M} D_{\textbf{Q};\mu} - \bar{N}_{\textbf{Q}}^\mu\bar{M} N_{\textbf{Q};\mu} \right) \nonumber\\
& &{} - \frac{1}{2} g_\pi\left( \bar{D}_{\textbf{Q}}^\mu S D_{\textbf{Q};\mu} - \bar{N}_{\textbf{Q}}^\mu S N_{\textbf{Q};\mu} \right)
+ \frac{1}{2} g_\pi\left( \bar{D}_{\textbf{Q}}^\mu \Phi N_{\textbf{Q};\mu} + \bar{N}_{\textbf{Q}}^\mu\Phi D_{\textbf{Q};\mu} \right) \nonumber\\
& &{} - \frac{g_A}{f_\pi}\left[ \bar{D}_{\textbf{Q}}^\mu\gamma_5\gamma_\nu \partial_\nu\Phi D_{\textbf{Q};\mu} - \bar{N}_{\textbf{Q}}^\mu\gamma_5\gamma_\nu\partial_\nu\Phi N_{\textbf{Q};\mu} + \bar{D}_{\textbf{Q}}^\mu\gamma_5\gamma_\nu \partial_\nu S N_{\textbf{Q};\mu} + \bar{N}_{\textbf{Q}}^\mu\gamma_5\gamma_\nu \partial_\nu S D_{\textbf{Q};\mu}\right] \ ,
\label{eq:LagDN}
\end{eqnarray}
where $S$ and $\Phi$ are defined as $M = S + i \Phi = S + 2 i \left( \pi^a T^a \right)$
with $\pi^a$ being the pion fields and $\mbox{tr} \left( T_a T_b \right) = (1/2) \delta^{ab}$.
\end{widetext}

As for the case of heavy mesons, the $\Delta$ term shifts the masses of the DHBs to the same direction, and $g_\pi$ term provides the mass difference between the chiral partners as
\begin{eqnarray}
\Delta M_{B;i} =
m_{D_{\textbf{Q},i}} - m_{N_{\textbf{Q},i}}  =  g_\pi\bar{M}_{ii},
\label{eq:MassDif}
\end{eqnarray}
which is exactly same as that for the heavy-light mesons in Eq,~(\ref{massdif:meson}). Then, the mass difference for the non-strange doubly heavy baryon is determined as
\begin{eqnarray}
m_{D_{\textbf{Q},q}} - m_{N_{\textbf{Q},q}} & = & 430~{\rm MeV}.
\end{eqnarray}
When we identify the quantum numbers of the $\Xi_{cc}^+$ observed in Ref.~\cite{Agashe:2014kda} as $\frac{1}{2}^+$ one can estimate the mass of the state $\Xi_{cc}^{\ast +}$ as $3950$~MeV.

We next consider the intermultiplet one-pion decays of the DHBs in the isospin symmetry limit. The relevant partial widths are expressed as
\begin{eqnarray}
\Gamma \left( \Xi_{cc}^{\ast ++} \to \Xi_{cc}^{++} + \pi^0 \right) & = & \Gamma \left( \Xi_{cc}^{\prime ++ \mu } \to \Xi_{cc}^{++ \mu } + \pi^0 \right) \nonumber\\
& = & \frac{(\Delta M_{B;u,d})^2}{8\pi f_\pi^2} \,\vert p_\pi \vert \ .
\end{eqnarray}
where $\vert p_\pi \vert $ is the three momentum of $\pi$ in the rest frame of the decaying DHB. Other partial widths of different possible charged states can be obtained by using the isospin relation. Our numerical results are given in Table.~\ref{tab:sum}.

\begin{table*}[htb]
\begin{tabular}{c|c|c|c}
\hline
\hline
\qquad Spectrum \qquad\qquad & \qquad Prediction (MeV) \qquad & \qquad Decay channel \qquad\qquad & \qquad Partial width (MeV) \qquad \\
\hline
$m_{\Xi_{cc}}^{\ast}$ & $ 3950 $ & $\Xi_{cc}^{\ast ++} \to \Xi_{cc}^{++} + \pi^0$ & $ 
331$  \\
\hline
$m_{\Xi_{cc}}^\mu$ & $ 3625 $ & $ \Xi_{cc}^{\ast ++} \to \Xi_{cc}^{+} + \pi^+$ & $ 
662$ \\
\hline
$m_{\Xi_{cc}}^{\prime \mu}$ & $ 4055 $ & $ \Xi_{cc}^{\prime ++} \to \Xi_{cc}^{++ \mu} + \pi^0$ & $ 
332 $ \\
\hline
$m_{\Omega_{cc}}^{\ast}$ & $ 4028 $ & $ \Xi_{cc}^{\prime ++} \to \Xi_{cc}^{+ \mu} + \pi^+$ & $ 
663 $ \\
\hline
$m_{\Omega_{cc}}^{\mu}$ & $ 3783 $ & $ \Omega_{cc}^{\ast +} \to \Omega_{cc}^{+} + \pi^0 $ & $ 
20 \times 10^{-3} $ \\
\hline
$m_{\Omega_{cc}}^{\prime \mu}$ & $ 4133 $ & $ \Omega_{cc}^{\prime + \mu} \to \Omega_{cc}^{+ \mu} + \pi^0$ & $ 
20 \times 10^{-3} $ \\
\hline
\hline
\end{tabular}
\caption{\label{tab:sum} Spectrum of the doubly charmed baryons and the partial widths of one-pion intermultiplet transitions. Here we take $m_{\Xi_{cc}} = 3520$~MeV~\cite{Agashe:2014kda}, $m_{\Omega_{cc}} = 3678$~MeV~\cite{Sun:2014aya} and $m_{\pi^\pm} = m_{\pi^0} \simeq 140$~MeV as input. Other partial widths of intermultiplet transitions can be obtained by using the isospin relation.
}
\end{table*}

We next consider the DHBs including a strange quark. In such a case, by using the spectrum of the heavy-light meson including a strange quark, one predicts~\cite{Bardeen:2003kt,Nowak:2003ra,Nowak:2004jg}
\begin{eqnarray}
m_{D_{\textbf{Q},s}} - m_{N_{\textbf{Q},s}} & = & m_{G_{s}} - m_{H_{s}} = 350~{\rm MeV}.
\end{eqnarray}
In this sector, due to the conservation of isospin, one might naively expect the dominant transition channel of $\Omega_{cc,s}^{\ast +}$ is $\Omega_{cc,s}^{\ast +} \to \Omega_{cc,s}^+ + \eta$. However, since the mass splitting $350~$MeV is smaller than the eta meson mass $m_\eta = 548~$MeV, this channel is forbidden due to the kinetic reason and dominant channel should be $\Omega_{cc,s}^{\ast +} \to \Omega_{cc,s}^+ + \pi^0$ arising from the $\eta$-$\pi^0$ mixing. The partial decay widths are expressed as
\begin{eqnarray}
\Gamma \left( \Omega_{cc}^{\ast +} \to \Omega_{cc}^+ + \pi^0 \right) & = & \Gamma \left( \Omega_{cc}^{\prime\mu + } \to \Omega_{cc}^{\mu +} + \pi^0 \right) \nonumber\\
& = &
\frac{ (\Delta M_{B;s})^2}{2\pi f_\pi^2} \Delta_{\pi^0\eta}^2\,\vert p_\pi \vert .
\end{eqnarray}
where
$\Delta_{\pi^0\eta} = -5.32 \times 10^{-3}$ is the magnitude of the $\eta$-$\pi^0$ mixing estimated in Ref.~\cite{Harada:2003kt}
based on the two-mixing angle scheme (see, e.g. Ref.~\cite{Harada:1995sj} and references therein). Since magnitude of the isospin breaking $\eta$-$\pi^0$ mixing is very small, the partial width of decay $\Omega_{cc,s}^{\ast +} \to \Omega_{cc,s}^+ + \pi^0$ is small. This situation is very similar to what happens in the heavy-light meson system in which the $D_{s0}(2317)$ is regarded as the chiral partner of $D_s$ and the dominant decay channel of the former is the isospin violating process $D_{s0}(2317) \to D_s + \pi^0$.

We further make a comment on the mass splitting of the baryons in a doublet which beyond the scope of the Lagrangian \eqref{eq:EffecL} we constructed. Here we just quote the result obtained in Ref.~\cite{Brambilla:2005yk},
\begin{eqnarray}
m_{\Psi_{QQ}^{(\prime)\mu}} - m_{\Psi_{QQ}^{(\ast)}} = \frac{3}{4} \left(  m_{D^{\ast}} - m_D \right) \simeq 105~{\rm MeV},
\label{eq:massdiffinter}
\end{eqnarray}
which is smaller than the pion mass. So that, in contrast to the heavy-light meson sector, the one-pion intramultiplet decays are forbidden in the DHB sector due to the kinetic reason. Note that it is reasonable to expect $60\%$ correction from $\mathcal{O}(1/m_c)$ to result \eqref{eq:massdiffinter} for doubly charmed baryons~\cite{Hu:2005gf} so that the one-pion decay channel could open for the intermultiplet decay.

Since the mass splitting between the chiral partners are from the spontaneous breaking of chiral symmetry, a particularly relevant problem is what will happen for this splitting in QCD under extreme condition. From the lessons in the heavy-light meson sector~\cite{Sasaki:2014asa,Suenaga:2014sga}, it is reasonable to expect that the magnitude of the mass spitting will be reduced in hot/dense matter. Such a scenario might be tested in a future scientific facility.

We finally want to stress that the present work mainly concerns the spectrum and dominant strong decay channels of the DHBs. Some other quantities such as the weak transitions of the DHBs through changing a heavy flavor are also interesting phenomenologically. These physics will be reported elsewhere.

In summary, 
we studied the spectrum and the dominant strong decay properties of the DHBs based on the chiral dynamics. We point out that the mass spitting between the lowest lying DHBs and their chiral partners is about $450$~MeV for the unflavored DHBs and $350$~MeV for the stranged DHBs. Moreover, we predicted that, due to the kinetic reason, the dominant decay channel of the parity odd strange DHB is isospin violating process therefore the partial width is small.


The work of Y.-L.~M. was supported in part by National Science Foundation of China (NSFC) under
Grant No.~11475071 and the Seeds Funding of Jilin University.
M.~H. was supported in part by the JSPS Grant-in-Aid for Scientific Research (S) No.~22224003 and (c) No.~24540266.


\begin{thebibliography}{99}

\bibitem{Moinester:1995fk}
For a review, see e.g.,
  M.~A.~Moinester,
  Z.\ Phys.\ A {\bf 355}, 349 (1996)
  [hep-ph/9506405].

\bibitem{Brodsky:2011zs}
  S.~J.~Brodsky, F.~K.~Guo, C.~Hanhart and U.~G.~Meissner,
  Phys.\ Lett.\ B {\bf 698}, 251 (2011)
  [arXiv:1101.1983 [hep-ph]].

\bibitem{Sun:2014aya}
  Z.~F.~Sun, Z.~W.~Liu, X.~Liu and S.~L.~Zhu,
  arXiv:1411.2117 [hep-ph].


\bibitem{Agashe:2014kda}
  K.~A.~Olive {\it et al.}  [Particle Data Group Collaboration],
  Chin.\ Phys.\ C {\bf 38}, 090001 (2014).

\bibitem{Nowak:1992um}
  M.~A.~Nowak, M.~Rho and I.~Zahed,
  Phys.\ Rev.\ D {\bf 48}, 4370 (1993)
  [hep-ph/9209272].

\bibitem{Bardeen:1993ae}
  W.~A.~Bardeen and C.~T.~Hill,
  Phys.\ Rev.\ D {\bf 49}, 409 (1994)
  [hep-ph/9304265].

\bibitem{Bardeen:2003kt}
  W.~A.~Bardeen, E.~J.~Eichten and C.~T.~Hill,
  Phys.\ Rev.\ D {\bf 68}, 054024 (2003)
  [hep-ph/0305049].


\bibitem{Nowak:2003ra}
  M.~A.~Nowak, M.~Rho and I.~Zahed,
  Acta Phys.\ Polon.\ B {\bf 35}, 2377 (2004)
  [hep-ph/0307102].

\bibitem{Nowak:2004jg}
  M.~A.~Nowak, M.~Praszalowicz, M.~Sadzikowski and J.~Wasiluk,
  Phys.\ Rev.\ D {\bf 70}, 031503 (2004)
  [hep-ph/0403184].

\bibitem{Harada:2012dm}
  M.~Harada and Y.~L.~Ma,
  Phys.\ Rev.\ D {\bf 87}, no. 5, 056007 (2013)
  [arXiv:1212.5079 [hep-ph]].

\bibitem{Falk:1991nq}
  A.~F.~Falk,
  Nucl.\ Phys.\ B {\bf 378}, 79 (1992).

\bibitem{Georgi:1990cx}
  H.~Georgi,
  Nucl.\ Phys.\ B {\bf 348}, 293 (1991).

\bibitem{Cho:1992gg}
  P.~L.~Cho,
  Phys.\ Lett.\ B {\bf 285}, 145 (1992)
  [hep-ph/9203225].


\bibitem{Savage:1990di}
  M.~J.~Savage and M.~B.~Wise,
  Phys.\ Lett.\ B {\bf 248}, 177 (1990).


\bibitem{Hu:2005gf}
  J.~Hu and T.~Mehen,
  Phys.\ Rev.\ D {\bf 73}, 054003 (2006)
  [hep-ph/0511321].

\bibitem{Harada:2012km}
  M.~Harada, H.~Hoshino and Y.~L.~Ma,
  Phys.\ Rev.\ D {\bf 85}, 114027 (2012)
  [arXiv:1203.3632 [hep-ph]].

\bibitem{Suenaga:2014sga}
  D.~Suenaga, B.~R.~He, Y.~L.~Ma and M.~Harada,
  Phys.\ Rev.\ D {\bf 91}, no. 3, 036001 (2015)
  [arXiv:1412.2462 [hep-ph]].

\bibitem{Harada:2003kt}
  M.~Harada, M.~Rho and C.~Sasaki,
  Phys.\ Rev.\ D {\bf 70}, 074002 (2004)
  [hep-ph/0312182].

\bibitem{Harada:1995sj}
  M.~Harada and J.~Schechter,
  Phys.\ Rev.\ D {\bf 54}, 3394 (1996)
  [hep-ph/9506473].




\bibitem{Brambilla:2005yk}
  N.~Brambilla, A.~Vairo and T.~Rosch,
  Phys.\ Rev.\ D {\bf 72}, 034021 (2005)
  [hep-ph/0506065].


\bibitem{Sasaki:2014asa}
  C.~Sasaki,
  Phys.\ Rev.\ D {\bf 90}, no. 11, 114007 (2014)
  [arXiv:1409.3420 [hep-ph]].


\end{thebibliography}
\end{document}